\documentclass[aps,prl,floatfix,twocolumn,showpacs,longbibliography]{revtex4-1}

\usepackage{verbatim}
\usepackage{dcolumn}
\usepackage{mathrsfs}
\usepackage{amsmath}
\usepackage{amssymb}
\usepackage{bm}
\usepackage{graphicx}
\usepackage{subfigure}
\usepackage{mciteplus}

\begin{document}


\title{Functional Subsystems and Quantum Redundancy in Photosynthetic Light Harvesting}

\author{Nolan Skochdopole}

\author{David A. Mazziotti}

\email{damazz@uchicago.edu}

\affiliation{Department of Chemistry and The James Franck Institute, The University of Chicago, Chicago, IL 60637}%

\date{Submitted August 24, 2011; Published in J. Phys. Chem. Lett. 2011, {\bf 2}, 2989}


\begin{abstract}

The Fenna-Matthews-Olson (FMO) antennae complex, responsible for
light harvesting in green sulfur bacteria, consists of three
monomers, each with seven chromophores.  Here we show that multiple
subsystems of the seven chromophores can transfer energy from either
chromophore~1 or~6 to the reaction center with an efficiency
matching or in many cases exceeding that of the full seven
chromophore system.  In the FMO complex these functional subsystems
support multiple quantum pathways for efficient energy transfer that
provide a built-in quantum redundancy.   There are many instances of
redundancy in nature, providing reliability and protection, and in
photosynthetic light harvesting this quantum redundancy provides
protection against the temporary or permanent loss of one or more
chromophores.  The complete characterization of functional
subsystems within the FMO complex offers a detailed map of the
energy flow within the FMO complex, which has potential applications
to the design of more efficient photovoltaic devices.

\end{abstract}

%
%
%
%
%
%

\maketitle

\section{Table of Contents Graphic}

After harvesting a photon of energy, the Fenna-Matthews-Olson (FMO)
antennae complex transfers it to the reaction center.

\begin{center}
\includegraphics[scale=0.25]{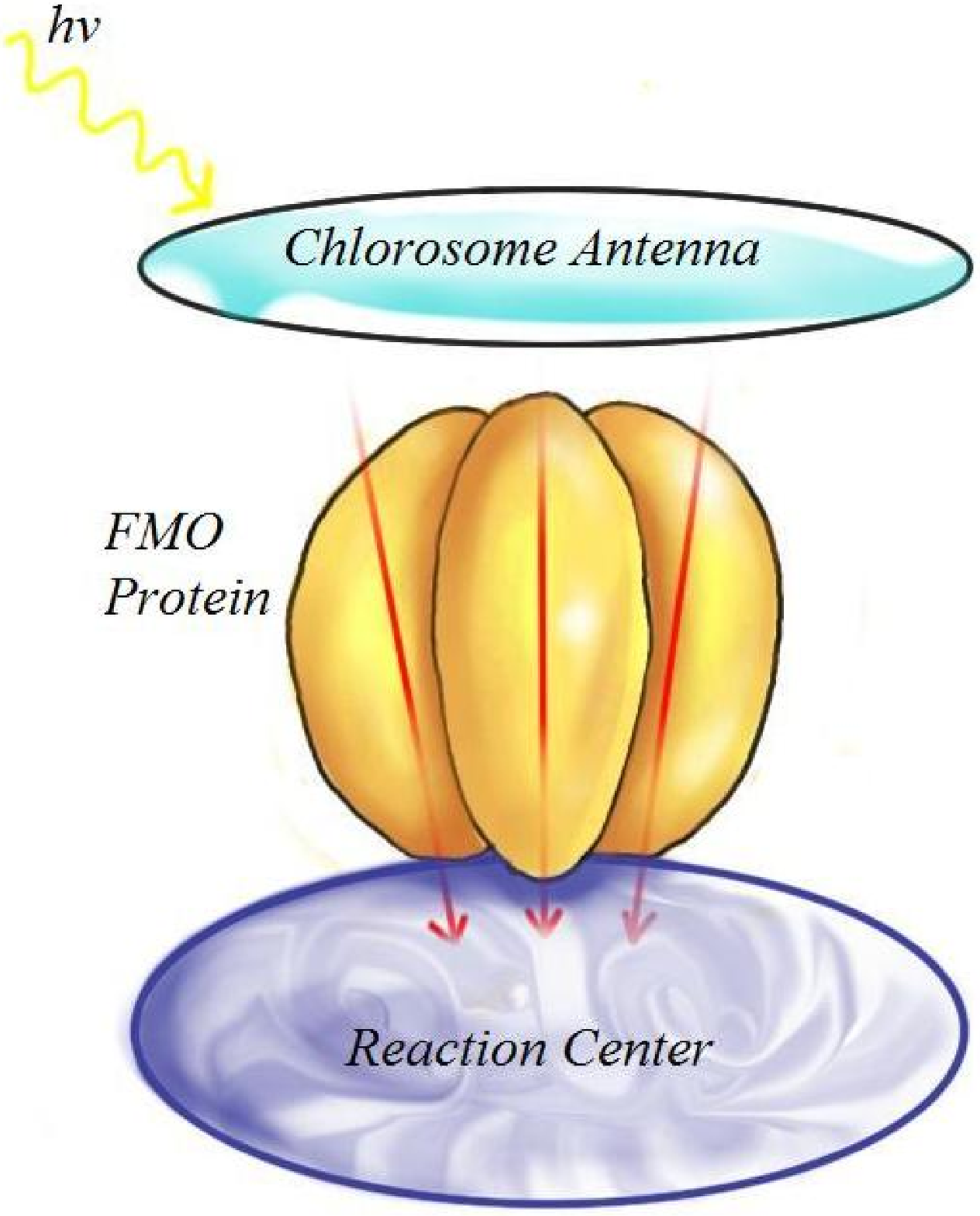}
\end{center}

{\bf Keywords}: functional subsystems, quantum redundancy,
chromophores, photosynthetic light harvesting, Fenna-Matthews-Olson
(FMO) antennae complex, entanglement, squared Euclidean distance

\section{Letter}

Most photosynthetic organisms have light-harvesting antennae
complexes composed of pigment molecules, called chromophores, whose
electrons are excited by sunlight~\cite{blankenship}.  The
excitation energy is transferred from the antennae complexes to
reaction centers with nearly a 100\% quantum
efficiency~\cite{sension}.  For many years this transfer was thought
to occur by a {\em classical} hopping mechanism, in which the
excitation energy transfers from one chromophore to another in a
downhill fashion until it reaches the reaction
center~\cite{excitons, scholes}, but recently, new experimental
evidence suggests that this transfer occurs by a {\em quantum}
coherent mechanism~\cite{flemingreview,BCR10}.  Quantum coherence in
the Fenna-Matthews-Olson (FMO) antennae complex of green sulfur
bacteria has been shown by two-dimensional Fourier-transform
electronic spectroscopy to last over 500~fs at ambient temperatures
and over 300~fs at physiological temperatures~\cite{engel,CWW10,
engel2}.

In this Letter we show that photosynthetic light harvesting exhibits
{\em quantum redundancy}.  By quantum redundancy we mean that
photosynthetic antennae complexes have multiple efficient quantum
pathways for transferring energy to the reaction center.  These
pathways involve different subsets of the antennae's chromophores,
which form {\em functional subsystems}.  The antennae complex known
as the FMO complex is a trimer where each monomer consists of seven
bacteriochlorophyll-a chromophores~\cite{fenna}.  Previous research
has shown that in each monomer initial excitation can occur on
either chromophore~1 or chromophore~6 and that this energy must move
to chromophore~3, which is most closely coupled with the reaction
center~\cite{renger}.  Here we show that within each monomer of the
FMO complex {\em there exist many subsets of the seven chromophores
with efficiencies close to or even better than the efficiency of the
entire set of seven chromophores}.  Quantum redundancy contributes a
robustness to nature's quantum device with potential survival
benefits.

The computations presented here reveal that the functional
subsystems achieve their efficiencies by a quantum mechanism similar
to that of the full system including the roles of entanglement and
environmental noise. We assess each subsystem's entanglement by a
global entanglement measure~\cite{HK05,K07b} based on the squared
Euclidean distance~\cite{PR02,BNT02,GM10,JM06,MH00,H78}.  The study
of these chromophore subsystems gives more information about the
role of each chromophore in the energy transfer in the whole FMO
complex~\cite{ZKR11,HC11,KR11,MBV11,SSH11,BWV10,L10,NT10,NG10,SIF10,
CDC10,WLS10}. This more in-depth understanding can provide insight
to other antennae complexes and ultimately be applied to create
synthetic solar cells that rival the efficiency of nature.

We consider a single monomer of the FMO complex with $M$ chromophore
sites with the Hamiltonian
\begin{equation}
\label{eq:H} {\hat H} =
\sum_{j=1}^{M}\hbar\omega_{j}\sigma_{j}^{+}\sigma_{j}^{-} + \sum_{j
\neq l} \hbar v_{j,l} (\sigma_{j}^{-}\sigma_{l}^{+} +
\sigma_{j}^{+}\sigma_{l}^{-}),
\end{equation}
where $\sigma_{j}^{+}$ ($\sigma_{j}^{-}$) creates (annihilates) a
single excitation on chromophore $j$.  The site energy of each
chromophore is $\hbar\omega_{j}$ while the coupling constant between
the pair of chromophores $j$ and $l$ is $\hbar v_{j,l}$.

The time evolution of the system's density matrix is governed by the
{\em quantum Liouville equation}
\begin{equation}
\frac{d}{dt} D = - \frac{i}{\hbar}[ {\hat H}, D ] + {\hat L}(D)
\end{equation}
where $D$ is the density matrix
\begin{equation}
D = \sum_{k,l}{ \rho^{k}_{l} | \Psi_{k} \rangle \langle \Psi_{l} |
},
\end{equation}
$\rho^{k}_{l}$ are the elements of the density matrix $D$ in the
basis set of wavefunctions $\{ \Psi_{k} \}$, and ${\hat L}$ is the
Lindblad operator, which accounts for interactions of the $M$
chromophores with the environment.  At $t=0$ we initialize the
density matrix with a single excitation (exciton) on either site~1
or 6. Because the Hamiltonian does not mix singly excited
wavefunctions with other wavefunctions, the set $\{ \Psi_{k} \}$
only includes $M$ wavefunctions formed from sequentially considering
a single excitation on each of the $M$ chromophores.

The Lindblad operator~\cite{L76,CDC10} is the sum of three operators
that account for dephasing, dissipation, and loss to the reaction
center (sink)
\begin{equation}
{\hat L}(D) = {\hat L}_{\rm deph}(D)+{\hat L}_{\rm diss}(D)+{\hat
L}_{\rm sink}(D)
\end{equation}
where
\begin{eqnarray}
{\hat L}_{\rm deph}(D) & = & \alpha \sum_{k}{2 \langle k | D | k
\rangle |k \rangle \langle k |  -\{| k \rangle \langle k |,D\} } ,
\\
{\hat L}_{\rm diss}(D) & = & \beta  \sum_{k}{ 2 \langle k | D | k
\rangle |g \rangle \langle g |  -\{| k \rangle \langle k |,D\} }
, \\
{\hat L}_{\rm sink}(D) & = & 2 \gamma \langle 3 | D | 3 \rangle | s
\rangle \langle s |  - \gamma \{| 3 \rangle \langle 3 |,D\} .
\end{eqnarray}
The $|g \rangle$ denotes the state in which each of the $M$
chromophores is in its ground state, the $|k \rangle$ represent the
$M$ excited states with each state having a single excitation on one
of the $M$ chromophores, the state $| s \rangle$ denotes the
reaction center (sink), and $| 3 \rangle$ indicates the excited
state of the third chromophore multiplied by the ground states of
the other $M-1$ chromophores.  The particular form of the Lindblad
operators chosen here follows the operators selected in previous
studies of the FMO complex~\cite{CDC10}.  Further refinements of the
system's interaction with the environment will not substantively
affect the general conclusions made here about the role of
functional subsystems within the FMO complex.  The model
incorporates two types of noise that affect the energy transfer.
First, dissipation transfers energy from the chromophores back into
the environment, wasting that energy. Second, dephasing dampens the
coherence within the chromophore system. Finally, the Lindblad
operator ${\hat L}_{\rm sink}(D)$ is responsible for transferring
energy from the third chromophore into reaction center, which acts
as an energy sink and becomes a $M+1$ site in the density matrix.
The rate parameters $\alpha$, $\beta$, and $\gamma$ were chosen in
atomic units to be $1.52 {\rm x} 10^{-4}$ , $7.26 {\rm x} 10^{-5}$,
and $1.21 {\rm x} 10^{-8}$, respectively.  These values are
consistent with those employed in Ref.~\cite{CDC10}.

{\em Entanglement} is a correlation that cannot occur in a classical
system; Bell~\cite{B64} defined entanglement as ``a correlation that
is stronger than any classical correlation.''  Parts of a quantum
system become {\em entangled} when the total density matrix for the
system cannot be expressed as a product of the density matrices for
the parts~\cite{HK05,K07b,HHH09}.  Global entanglement~\cite{HK05,
K07b,HHH09} in the system is measured by the {\em squared Euclidean
distance} between the density matrix $\rho$ and the nearest {\em
classical} density matrix $\xi$~\cite{PR02,BNT02,GM10,JM06,MH00,H78}
\begin{equation}
\sigma(\rho) = \sum_{i,j}{ |\rho^{i}_{j} - \xi^{i}_{j}|^2 } .
\end{equation}
In the basis of $M$ wavefunctions, in which each wavefunction has a
single excitation on one of the $M$ chromophores, this measure
corresponds to the sum of the squares of the off-diagonal elements
of $\rho$, or the sum of the squares of the concurrences.  The
squared Euclidean distance is zero if and only if the excitations
(excitons) on the chromophores are not entanglement.  Both squared
Euclidean distance and {\em quantum relative
entropy}~\cite{HHH09,SIF10}, another measure of global entanglement,
are Bergmann distances.

\begin{figure}[ht!]

\label{fig:density}

\begin{center}
\subfigure{\includegraphics[scale=0.3]{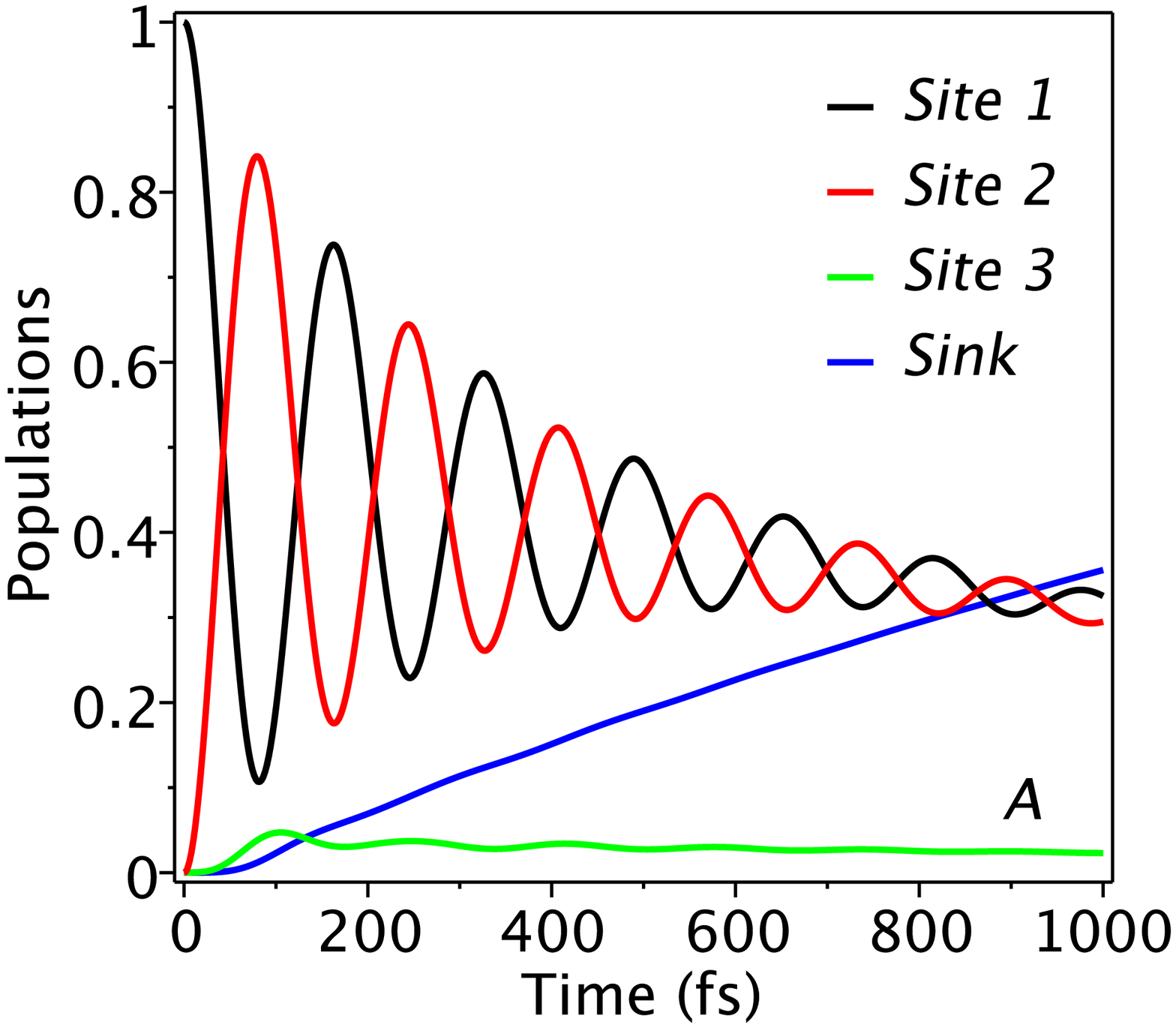}}
\subfigure{\includegraphics[scale=0.3]{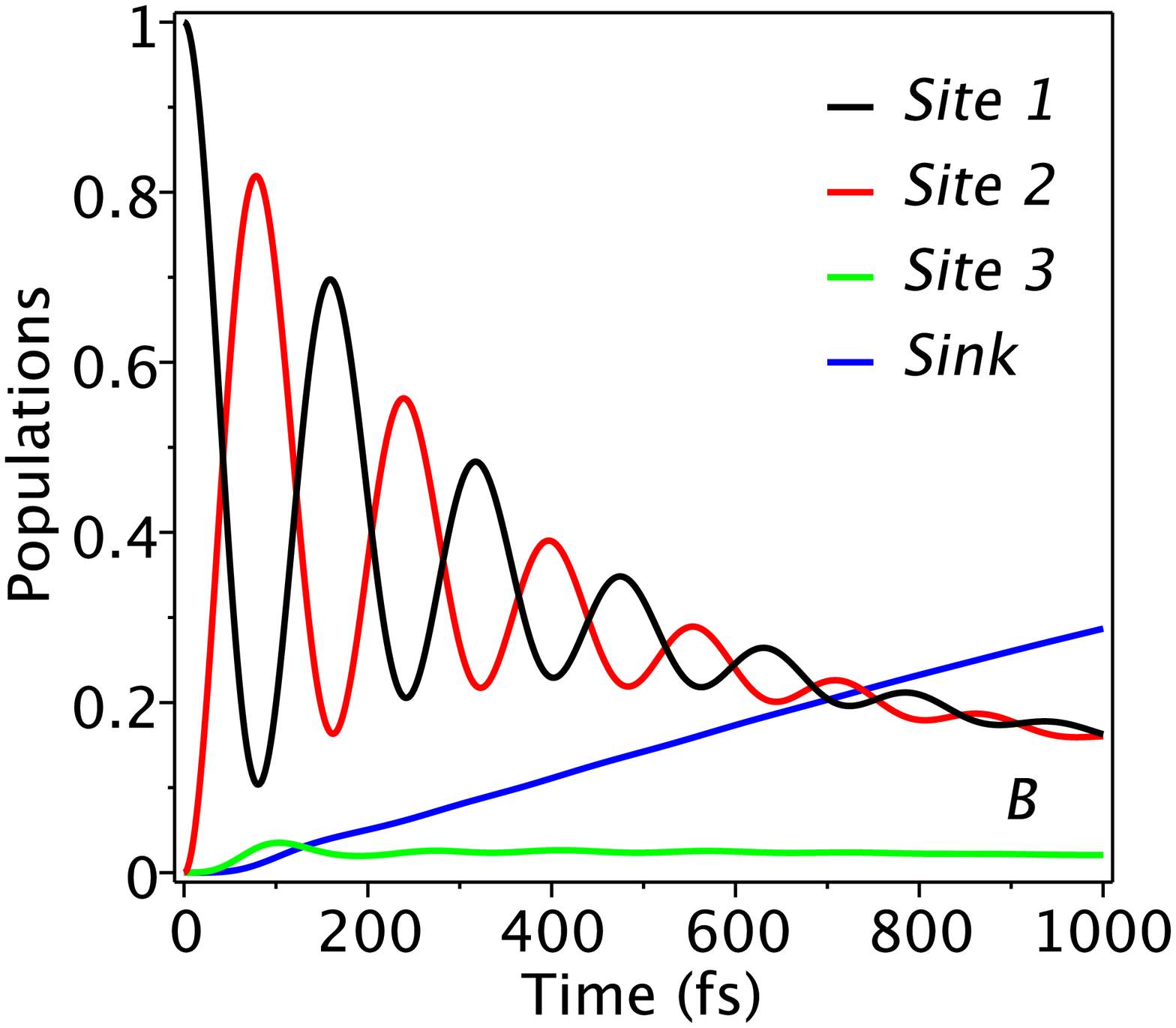}}
\end{center}

\caption{Populations of chromophores 1-3 and the reaction center for
(A) the subsystem 123 and (B) the full FMO system 1234567. Subsystem
123 of the FMO complex transfers energy more efficiently from
chromophore~1 to the reaction center (sink) than the full FMO
system.}

\end{figure}

In Figs.~1-3 and Table I we study subsystems of the FMO complex, in
which certain of the seven chromophores are ``turned off'' and the
energy transfer is tracked within a smaller set of chromophores. For
the full system of seven chromophores we employ the 7x7 Hamiltonian
given in Ref.~\cite{CDC10}; chromophores are removed from the full
system by deleting rows and columns of the Hamiltonian matrix.  We
denote the subsystem by the numbers of the chromophores retained;
for example, the subsystem with chromophores~1, 2, and 3 is denoted
as 123. Figure~1 shows the exciton populations of chromophores 1, 2,
3, and the reaction center for reduced system 123 and the full
system. We observe that the two systems are quite similar, and yet
they exhibit some important differences.  The reduced system
features a slightly slower decay of the exciton populations in
chromophores~1 and~2, and most significantly, it transfers energy
from chromophore~1 to the reaction center more quickly than the full
system.  Because chromophores 4-7 draw some of the excitation energy
in the full system, the populations of chromophores 1-3 and the
reaction center are lower in the full system than in the reduced
system.

The efficiency of energy transfer from site~1 to the sink is
compared in Fig.~2 for several subsystems, 123, 1234, 12345 (not
shown because its efficiency is similar to 1234), and 123456, as
well as the full system.  We observe that subsystems 123, 1234, and
12345 are more efficient than the full system even though efficiency
decreases as we add chromophores from 123 to 123456. The addition of
chromophores 4, 5, and 6 increases the number of sites for the
energy to enter and thereby decreases the population of every site,
including the reaction center.  The addition of chromophore~7 to
123456, however, improves efficiency; because chromophore~7 has a
lower energy than chromophores~5 and 6, it offers a new downhill
quantum path through chromophore~4, even lower in energy, to
chromophore~3 and the reaction center.  The importance of
chromophore~7 can also be seen from Table~I, which shows that the
subsystem containing chromophores 12347 is the most efficient
system, again because of the downhill path created from chromophore
7 to the reaction center.

\begin{figure}[ht!]

\label{fig:sink}

\begin{center}
\includegraphics[scale=0.3]{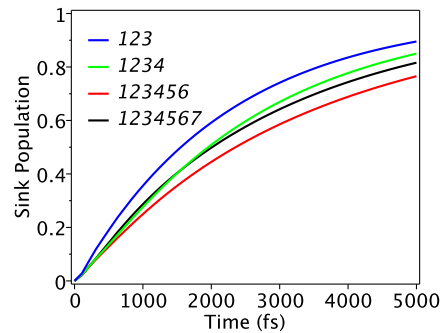}
\end{center}

 \caption{Population in the reaction center for several
chromophore subsystems.  The subsystems 123, 1234, and 12345 (not
shown) are more efficient than the full FMO system at transferring
energy from chromophore~1 to the reaction center (sink).}

\end{figure}

Table~I reports the efficiencies of subsystems with the initial
excitation on either chromophore~1 or chromophore~6.  Subsystems
with both chromophores~1 and 6 are included in both categories.  The
table shows that most of these subsystems are more efficient than
the full system at transferring energy from either chromophore~1
{\em or} chromophore~6 to the reaction center.  However, only one
subsystem 123467 is more efficient than the full system when
efficiency is assessed with equal weight given to the excitation
pathways from chromophores~1 {\em and} 6 to the reaction center.
Although this model suggests that chromophore~5 is superfluous, it
may have other necessary duties in nature that are not considered in
the present model.  The superfluity of chromophore~5 is predictable
from the Hamiltonian of the system, as the site energy for
chromophore 5 is much higher than the other
chromophores~\cite{renger}.  However, it is worth noting that this
is only the case in the FMO complex for {\em Prosthecochloris
aestuarii }, which we are studying in this Letter, while in the FMO
complex for {\em Chlorobium tepidum}, chromophore~5 is not the
highest in energy and its exclusion would most likely make the
system less efficient.  Very recent crystallographic and quantum
chemistry studies~\cite{SMM11,TWG09} indicate that there is likely
an eighth chromophore in the FMO of {\em Prosthecochloris
aestuarii}, which due to sample preparation is not present in the
ultrafast spectroscopic studies~\cite{engel,CWW10,engel2}.  An
eighth chromophore would further support the existence and
importance of functional subsystems with the seven chromophores in
recent ultrafast experiments serving as a functional subsystem of
the full system in nature.

\begin{table}
\caption{Populations of the reaction center of different chromophore
subsystems at 2~ps.}
\begin{ruledtabular}
\begin{tabular}{cccc}
& {} & \multicolumn{2}{c}{Reaction Center Population} \\
\cline{3-4}
Initial Excitation & System & No Dephasing & Dephasing\\
\hline
Chromophore 1 & 123 & 0.387 & 0.592 \\
{} & 1234 & 0.161 & 0.509 \\
{} & 12345 & 0.144 & 0.506  \\
{} & 123456 & 0.120 & 0.445 \\
{} & 1234567 & 0.459 & 0.498 \\
{} & 12347 & 0.347 & 0.564 \\
{} & 123467 & 0.314 & 0.514 \\
Chromophore 6 & 346 & 0.010 & 0.090 \\
{} & 367 & 0.009 & 0.026 \\
{} & 3467 & 0.048 & 0.486 \\
{} & 3456 & 0.045 & 0.305 \\
{} & 34567 & 0.103 & 0.445 \\
{} & 234567 & 0.086 & 0.434 \\
{} & 1234567 & 0.139 & 0.433 \\
{} & 123467 & 0.074 & 0.456 \\
\end{tabular}
\end{ruledtabular}
\end{table}

The subsystems, we observe, function by a quantum mechanism similar
to that observed in the full system. Two features of this mechanism,
which have received significant attention for the full system in the
literature~\cite{ZKR11,HC11,KR11,MBV11,SSH11,BWV10,L10,NT10,NG10,SIF10,
CDC10,WLS10}, are: (i) dephasing from environmental noise and (ii)
entanglement of excitons.  As in the full system, the dephasing
assists transport of the excitons from chromophore~1 or 6 to the
reaction center by facilitating the distribution of the exciton
population from its initial chromophore.  The efficiencies of
different subsystems with and without dephasing are reported in
Table~I.  The perceived enhancement is a characteristic of systems
of heterogeneous chromophores, that is chromophores with different
site energies. We also observe in Fig.~3, displaying the
entanglement in several subsystems as well as the full system, that
the entanglement of excitons exhibits similar behavior in the
reduced systems as in the full system.  We compute the entanglement
to demonstrate that the transfer of energy for both reduced and full
systems {\em within the model} occurs by the same quantum mechanism.
(It is not the purpose of the present Letter to examine the
relationship between entanglement and efficiency, which has been
explored in previous work including a recent paper by one of the
authors~\cite{M11b}.)  The addition of each chromophore slightly
increases the total entanglement of the system, as measured by the
squared Euclidean distance, by adding extra off-diagonal elements to
the density matrix.  Each system shows that the entanglement lasts
for a few hundred femtoseconds, which matches previous experimental
data as well as predictions from other theoretical
models~\cite{engel2,ishizaki}.

\begin{figure}[ht!]

\label{fig:ent}

\begin{center}
\includegraphics[scale=0.3]{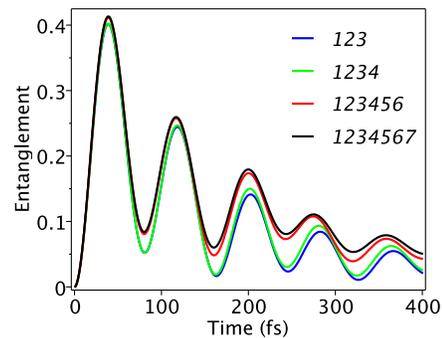}
\end{center}

\caption{Entanglement of excitons for several chromophore subsystems
as a function of time.  A measure of global entanglement is shown as
a function of time (fs) for several chromophore subsystems, 123,
1234, and 12346, as well as the full system 1234567.  Like the full
system, the subsystems exhibit entanglement of excitons between the
chromophores.}

\end{figure}

Efficient photosynthetic light harvesting, according to recent
experimental evidence, likely occurs by a quantum rather than a
classical mechanism.  In this Letter we have identified multiple
functional subsystems of the FMO light-harvesting complex whose
efficiencies are comparable to the efficiency of the full FMO
complex.  Many of these subsystems are, in fact, more efficient in
transferring energy from either chromophore~1 or~6 to the reaction
center, and one of these subsystems is even more efficient in
transferring energy from both chromophores~1 and 6 to the reaction
center.  These functional subsystems provide evidence for quantum
redundancy in photosynthetic light harvesting.  There exists
multiple quantum pathways or circuits that are capable of
transferring energy efficiently to the reaction center.  For
example, subsystem 123 transfers energy from chromophore~1 to the
reaction center with 59.2\% efficiency (after 2~ps), and subsystem
3467 transfers energy from chromophore~6 to the reaction center with
48.6\% efficiency.  This built-in redundancy likely provides
benefits to the natural system because damage to any chromophore
save 3 will not disrupt light harvesting.  The characterization of
functional subsystems within the FMO complex offers a detailed map
of the energy flow within the FMO complex with potential
applications to the design of more efficient photovoltaic devices.

\begin{acknowledgments}

DAM gratefully acknowledges the National Science Foundation, the
Army Research Office, Microsoft Corporation, the Henry and Camille
Dreyfus Foundation, and the David and Lucile Packard Foundation for
their generous support.

\end{acknowledgments}



\end{document}